\documentclass[12pt]{article}
\usepackage{graphicx}
\newcommand{\beq}{\begin{equation}}
\newcommand{\eeq}{\end{equation}}
\newcommand{\bea}{\begin{eqnarray}}
\newcommand{\eea}{\end{eqnarray}}

\begin{document}
\begin{titlepage}
\begin{flushleft}
       \hfill                      {\tt hep-th/0103071}\\
       \hfill                       FIT HE - 01-01 \\
       \hfill                       Kagoshima HE-01-2 \\
\end{flushleft}
\vspace*{3mm}
\begin{center}
{\bf\LARGE Stability of Randall-Sundrum brane-world \\ }
\vspace*{5mm}
{\bf\LARGE and tachyonic scalar \\ }
\vspace*{12mm}
{\large Kazuo Ghoroku\footnote{\tt gouroku@dontaku.fit.ac.jp} and
Akihiro Nakamura\footnote{\tt nakamura@sci.kagoshima-u.ac.jp}}\\
\vspace*{2mm}

\vspace*{2mm}

\vspace*{4mm}
{\large ${}^1$Fukuoka Institute of Technology, Wajiro, Higashi-ku}\\
{\large Fukuoka 811-0295, Japan\\}
%\vspace*{10mm}
\vspace*{4mm}
{\large ${}^2$Kagoshima University, Korimoto 1-21-35}\\
{\large Kagoshima 890-0065, Japan\\}
\vspace*{10mm}
\end{center}

\begin{abstract}
We address the behaviour of the scalar field with negative mass squared
in the five dimensional AdS space where the Randall-Sundrum brane-world is 
embedded. We point out that the tachyonic scalar allowed in the bulk space
destabilizes the embedded brane-world where the cosmological constant is zero. 
The resolution of this instability is discussed from the viewpoint of
AdS/CFT correspondence.

\end{abstract}
\end{titlepage}

\section{Introduction}

People believe that our four dimensional world would be obtained
from the ten or eleven dimensional superstring theory. The D-brane
solutions have attracted large attentions of many people, and an
interesting geometry has been obtained from the D3-brane of type IIB theory.
Near the horizon of the D3-branes, 
we find the configuration of $AdS_5\times S^5$. The string
theory under this background would describe
the four dimensional conformal symmetric Yang-Mills theory on the boundary
of $AdS_5$ \cite{M1,GKP1,W1,Poly1} where flat Minkowski space is realized. 

On the other hand, a thin three-brane (RS brane)
can be embedded in $AdS_5$ space ~\cite{RS1,RS2} at any
point of the transverse coordinate, which is interpreted as the
energy scale of the conformal field theory on the boundary. 
A promising picture is to consider this three-brane as our four-dimensional
world. This picture gives an alternative to the standard Kaluza-Klein (KK)
compactification via the localization of the zero mode of the gravitation 
\cite{RS2}.
It could also give a new
explanation for the problems of the hierarchy between Planck mass and
the weak-electromagnetic mass scales. This proposal is distinct from
the one given in \cite{ADD,AADD,Ant}.
Further, it opens a new
possible explanation of 
the smallness of the four dimensional cosmological constant
in our world without fine-tuning \cite{ArHa,KaSc}. 

In this framework, gravity is defined in the five dimension, but its
zero mode is trapped in the brane, then we can observe the
usual 4-dim Newton law in the brane-world. 
Other massive KK modes living in the five dimension
are observed as the correction to the Newton law. 
From the viewpoint of the string theory,
the bulk continuum modes are described also by the idea of the
AdS/CFT correspondence \cite{11,DL}. 
The correction to the Newton law can be obtained
as the quantum correction from the CFT coupled with the gravitation \cite{DL}.

Other than the graviton, there would be many kinds of fields 
in the bulk to be studied since the five dimensional
bulk theory would be obtained by the reduction of the ten dimensional IIB
theory or eleven dimensional M theory. In any case, we would obtain
many number of scalar fields in the reduced five dimensional theory 
\cite{PPN,BC}.
Many people have tried to derive the domain wall solution, which could 
be identified with the RS brane in a thin limit, from the reduced supergravity
\cite{Be,DuLi}.
In these approaches, the scalar 
fields play an important role, so it would be important to know their behaviour
in constructing the RS brane(s) in the five
dimensional space.

The purpose of this paper is to
study the scalar with its wide range of mass allowed
in the $AdS_5$. In general, negative mass 
squared is allowed and it is bounded from below 
$M^2\ge -4/L^2$ where $L$ is the radius of AdS space \cite{Bre}. 
Our interest is to see how this scalar field can be observed
from the brane world
embedded in $AdS_5$. For $M=0$ case, it is well-known that the zero mode
of this scalar is localized in the brane and KK modes are observed as the
correction to the zero mode propagator.
We extended the
analysis to the general case of the scalar mass $M$ according to the technique
used in the linearized gravity \cite{11}. In the next section, the solutions of
the wave equation
are examined to see the localization of some eigenstate of the four
dimensional mass of the scalar. In the section three, the propagator
for this scalar
in the bulk space is given by considering appropriate boundary conditions
to see the effective two point function of this scalar in the brane-world.  
We can see that the two different approaches lead to the consistent results.
The most important result is that a tachyonic localized-state appears in the
case of bulk tachyonic-scalar, $M^2<0$.
In section four, an idea for the resolution of this problem is discussed
according to the idea of the AdS/CFT correspondence. 
In the section five, the case of two branes is examined and we find the scalar
considered here has nothing to do with the stability of the distance between
the two branes. The concluding remarks are given in the final section.

\section{Localized state of the scalar }

Here we start from an effective five-dimensional action which is
responsible for the construction of the five dimensional
background. It is given in the Einstein frame as,
\beq
    S_5 = {1\over 2\kappa^2}\Bigg\{
      \int dX^5\sqrt{-G} (R + \Lambda + \cdots)
          +2\int dx^4\sqrt{-g}K\Bigg\}, \label{action}
\eeq
where dots denote the contents other than the gravitation, and
$K$ is the extrinsic curvature on the boundary.
The brane action,
\beq
    S_{\rm b1} = -{\tau\over 2}\int dx^4\sqrt{-g}, \label{baction}
\eeq
is added to $S_5$ and we obtain the 5-dimensional AdS background
\beq
 ds^2= e^{-2|y|/L}\eta_{\mu\nu}dx^{\mu}dx^{\nu}
           +dy^2  \, \label{metrica}
\eeq
where $\tau=6/(L\kappa^2)$ and
$L=\sqrt{6/\kappa^2\Lambda}$ which denotes the radius
of AdS space. The coordinates parallel to the brane are denoted by $x^{\mu}$
and $y$ is the coordinate transverse to the brane. 

The fields represented by dots
in the bulk action are not needed to construct above background $AdS_5$
with RS brane. They are however necessary in solving
the problem of the cosmological constant without fine-tuning or in constructing
the domain wall solution as a thick
brane. Here we concentrate our attention on the problem of the stability
of the brane-world when we consider the bulk scalar 
which does not play any role in
constructing the background and it is a part of $\cdots$ in (\ref{action}).
We consider the scalar with mass $M$, and it is denoted by $\Phi$. 
Then its field equation in the AdS
background is given as
\beq
 {1\over \sqrt{-G}}\partial_A(\sqrt{-G}G^{AB}\partial_B\Phi)
        -M^2\Phi=0, \label{scalar}
\eeq
where $A, B$ denote five dimensional suffices. This equation can be
rewritten as a one dimensional eigenvalue equation similar to the
Schr\"odinger equation as follows. By decomposing as 
$\Phi=\phi(x)\hat{\psi}(y)e^{3|y|/2L}$ and assuming 
$\eta^{\mu\nu}\partial_{\mu}\partial_{\nu}\phi=m^2\phi$, we
obtain
\beq
 [-\partial_z^2+V(z)]\hat{\psi}(z)=m^2\hat{\psi}(z) , \ \label{scal2}
\eeq
where $z={\rm sgn}(y)L(e^{|y|/L}-1)$ and
\beq
 V(z)={a \over (|z|/L+1)^2}-{3\over L}\delta(z), \qquad 
a={15\over 4L^2}+M^2 . \label{poten}
\eeq

Equation (\ref{scal2}) is regarded as the one dimensional Schr\"odinger
equation with the energy eigenvalue $m^2$. For the case of $M=0$, this
is equivalent to the one of the graviton. In this case, $a=15/4L^2>0$ and
the potential has the shape of a volcano (see Fig.~1). 
This is a necessary condition
to localize the zero mode ($m=0$) in the brane.

\begin{figure}
\begin{center}
\includegraphics[width=12truecm]{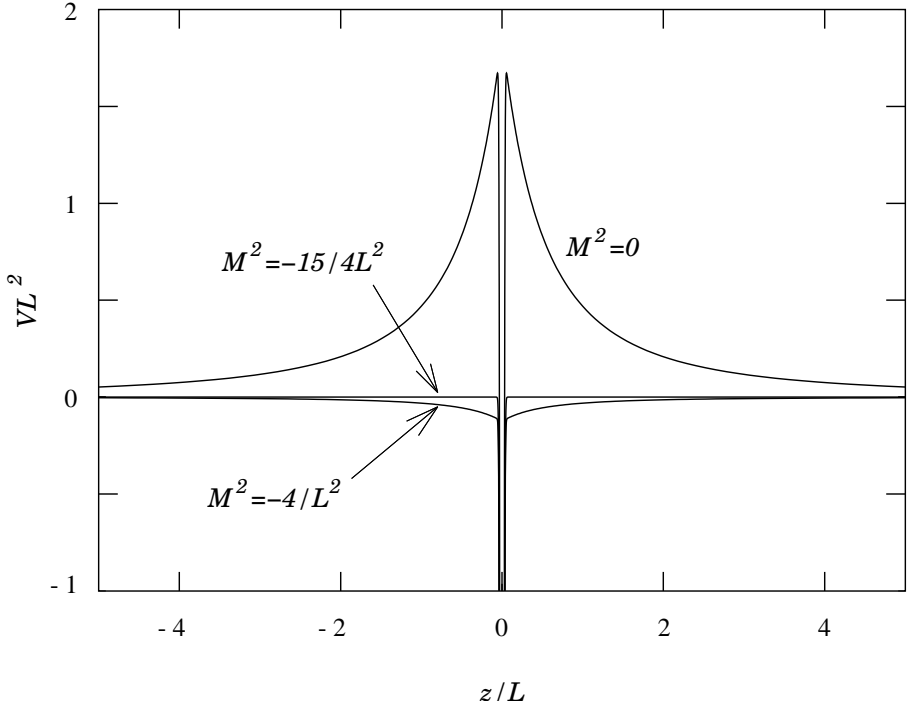}
\caption[]{Shape of the potential $V(z)$}
\end{center}
\end{figure}

In general, the shape of the potential varies with the scalar mass $M$, which
is bounded as $M^2\geq -4/L^2$ in the AdS background by the positivity
of the energy. Within this bound, 
we can see that the volcano type potential 
disappears at $M^2=-15/4L^2$, where $a=0$.
So we expect a tachyonic bound state in the brane
at least in the region $-15/4L^2>M^2\geq -4/L^2$. 
This implies that the RS brane-world with flat background would be
unstable if the above tachyonic scalar resides in the bulk.

The eigenvalue of $m^2$ for the tachyonic bound state is obtained as follows.
Firstly, $m^2$ is replaced by $-m^2$ in (\ref{scal2}) since we are solving
this equation in the region $m^2<0$. After this replacing, $m$ is considered
as a real number. Secondly,
we demands that the wave function of the bound state would be 
normalizable in the sense that the following integral with respect to $y$
is finite,
\beq
 \int dy\sqrt{-G}(G^{AB}\partial_A\Phi\partial_B\Phi + M^2\Phi^2)
      =\sqrt{-g}
    (g^{\mu\nu}\partial_{\mu}\tilde{\phi}\partial_{\nu}\tilde{\phi}
            + m^2\tilde{\phi}^2) , \label{norm1}
\eeq
where the right hand side provides the effective action for the 
bound state scalar, which is denoted by $\tilde{\phi}(x)$.
It is related to $\phi(x)$ defined previously by
\beq
 \tilde{\phi}(x)=\sqrt{\int_0^{\infty}dy\hat{\psi}^2e^{|y|/L}}
               \phi(x).  \label{norm}
\eeq
Then the condition of the normalizability is equivalent to the finiteness
of the integral in (\ref{norm}).
In general, the solution of (\ref{scal2}) can be
written by the linear combination of two kinds of modified Bessel functions.
But the solution here is written by one of them due to the requirement
of the normalizability,
\beq
  \hat{\psi}=Nx^{1/2}K_{\nu}(mx), \label{tachw}
\eeq
where $x=|z|+L$, $\nu=\sqrt{4+M^2L^2}$ and $K_{\nu}$ denotes the second
kind of modified Bessel function, which converges at large $mx$.{\footnote
{Here we notice that (\ref{tachw}) is obtained as the solution of
(\ref{scal2}) where $m$ is replaced by $im$.}}
$N$ is a
normalization constant. 

We need the following boundary condition for this solution at the position 
of the brane, at $z=0$, due to the $\delta$-function potential,
\beq
 \partial_z\hat{\psi}|_{z=0}=-{3\over 2L}\hat{\psi}|_{z=0}. \label{bound}
\eeq
This condition leads to the following,
\beq
    (2+\nu)K_\nu(mL)=(mL)K_{\nu+1}(mL). \label{bound2}
\eeq
This  equation determines the eigenvalue of $m$.

From the discussion given above,
the region to be examined would be restricted to the region
$-15/4L^2>M^2\geq -4/L^2$ or equivalently $1/2>\nu\geq 0$ where
the volcano type of shape of the potential disappears. However we
solve the above equation (\ref{bound2}) by extending
the region to $\nu\geq 0$. After a numerical research, (i) 
we can't find any solution for $\nu>2$ 
expectedly. (ii) At $\nu=2$ ($M=0$),
we find the solution at $m=0$, and this reflects the localization of the 
zero mode of a massless scalar in the bulk. This is parallel to the
localization of the zero mode of the graviton.
(iii) In the region $2>\nu>1/2$, the potential
preserves the volcano shape as in the case of $\nu=2$, so one would expect 
to find the localization of the zero mode. However, we couldn't find it and
the localization of a tachyonic state was found instead.
(iv) For $1/2\geq\nu\geq 0$, the localization of a tachyonic state was found
expectedly.

As a result, we find only one solution
for $2\geq\nu\geq 0$, and there is no solution for $\nu>2$. 
The value of $m$ for the localized state changes continuously with $\nu$
as shown in Fig.~2.
\begin{figure}
\begin{center}
\includegraphics[width=12truecm]{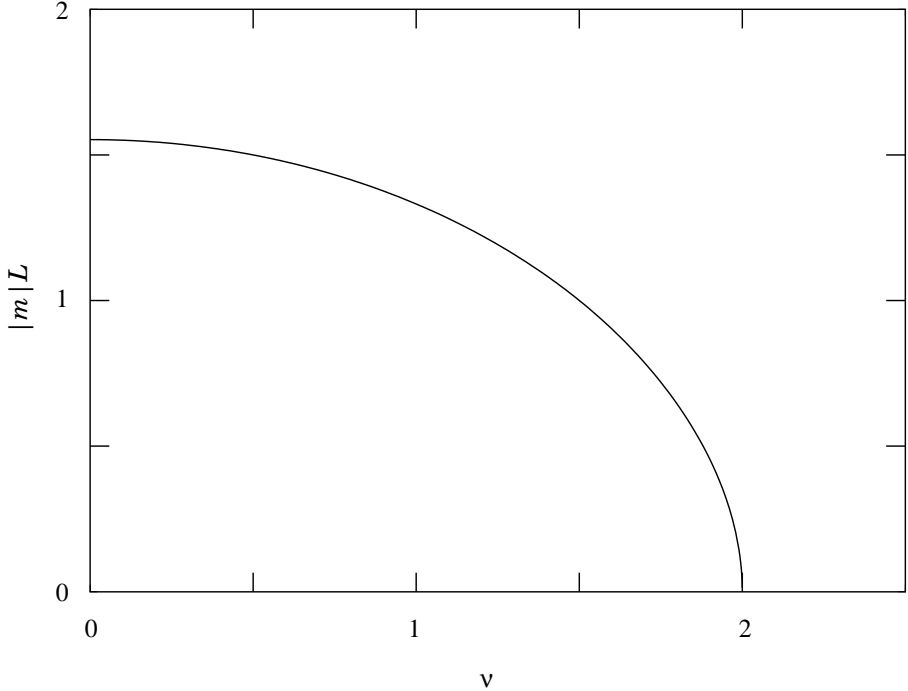}
\caption[]{Location of the tachyon pole vs $\nu$}
\end{center}
\end{figure}
Namely, $mL$ smoothly
increases from $0$ to about $1.5$ as $\nu$ decreases
from $2$ to $0$. This seems to be curious, but we can reassure 
this point in the next section 
from the analysis of the five dimensional propagator of the scalar.
The existence of only one solution for each $\nu$
is consistent with the $\delta$-functional attractive 
potential in the eigenvalue equation. The localization of zero mode
is realized only for the bulk massless field. The tachyonic localized-state is
always
seen for the case of the tachyonic bulk scalar, $M^2<0$. This can be 
interpreted as
the penetration of the tachyonic state from the bulk into the brane.

%%%%%%%%%%%%%%%%%%   Scalar propagator   %%%%%%%%%%%%%%%%%%%%%%%%%
\section{Scalar propagator}

According to Ref.~\cite{11}, the Green function of the scalar considered in the
previous section is 
examined in the $d$+1 dimensional AdS space to see its effective propagator
observed in $d$ dimensional flat space of the brane.
After obtaining the Green function, we go back to the case of $d=4$.
Here we work in the brane background of the following
AdS metric, 
\beq
    ds^2 = {{L^2\over{z^2}}({dz^2} + {dx_d^2})}  .
\eeq
We can consider that the brane is located at $z=L$, where $y/L=\ln(z/L)$.  
The scalar Green function $\Delta_{d+1} (X,X^{\prime})$ is defined as
\beq
 (\nabla^2 - M^2)\Delta_{d+1} (X,X^{\prime}) = 
    {\delta^{d+1}(X - X^{\prime})\over {\sqrt {-G}}}\ , \label{sceq}
\eeq
with the Neumann boundary condition,
\beq
\partial_z \Delta_{d+1} (X,X')|_{z=L} = 0 \ .
\eeq
This condition is consistent with the orbifold boundary conditions
at the brane.

Equation (\ref{sceq}) is rewritten by an ordinary differential equation
via following Fourier transform in the $d$ dimensions along the brane,
\beq
 \Delta_{d+1}(x,z;x',z') = \int {d^d p \over 
(2\pi)^d}e^{ip(x-x')} \Delta_p(z,z'),   \label{fourier}
\eeq
and the redefinition of the Fourier component, 
$\Delta_p = ({zz' \over L^2})^{d\over 2}{\hat \Delta}_p$ .
Then the equation for ${\hat \Delta}_p$ is obtained as
\beq
  \left(z^2\partial_z^2 + z\partial_z +q^2z^2 - \nu^2\right) 
\hat{\Delta}_p(z,z')=Lz\delta(z-z').   \label{hatlapl}
\eeq
where
\beq
  q=\sqrt{-p^2},\qquad \nu=\sqrt{(d/2)^2+L^2M^2}  . \label{nu}
\eeq
%%%%%%%%%%%%%%%

After this,
the procedure to obtain the propagator is parallel 
to \cite{11}. A case that is of particular interest here is that
where the arguments of $\Delta_{d+1}$ is on the brane, 
$z=z'=L$. In this case, the propagator is expressed as
\beq
  \Delta_{d+1}(x,L;x',L) = \int
{d^dp \over (2\pi)^d} e^{ip(x-x')} {1\over F(q,L,M)} \ ,  \label{braneGF}
\eeq
\beq
    F(q,L,M)= {d-2\nu\over 2L}  + q{H_{\nu-1}^{(1)}(qL) \over 
                                      H_{\nu}^{(1)}(qL)} . \label{prop}
\eeq
From this result, some interesting features are observed. Hereafter
we consider the case of $d=4$.

(i) First, this is reduced to the form given in \cite{11} for 
$M=0$ (or $\nu=d/2$), and
$\Delta_{d+1}(x,L;x',L)$ is separated to the $d$-dimensional massless
propagator and the exchange of the Kaluza-Klein states. 
\beq
 \Delta_{d+1}(x,L;x',L) = {d-2\over L} \Delta_d(x,x') +
\Delta_{KK}(x,x')\ .   \label{propsep}
\eeq
Here
\beq
\partial_{\mu}\partial^{\mu}\Delta_{d}(x,x')=\delta^{d}(x-x')\  ,
\label{Gddef}
\eeq
\beq
 \Delta_{KK}(x,x') = -\int
{d^dp \over  (2\pi)^d} e^{ip(x-x')}{1\over q}  
{H_{{d \over 2}-2}^{(1)}(qL) \over 
 H_{{d \over 2}-1}^{(1)}(qL)}\ .  \label{kkcontrib}
\eeq
$\Delta_{d}(x,x')$ represents the localized massless state
in the $d(=4)$ dimensional brane. The leading part of the 
summation of the KK exchanges gives $1/r^3$ potential as in the case of the
gravity.

(ii) For $\nu>2$, we can see the massive pole behaviour by expanding 
$F(q,L,M)$ near small $qL$,
\beq
  (d/2-\nu)/L + (1/2)(L/(\nu-1))q^2 ,  \label{appro}
\eeq
which implies a pole at $(qL)^2=2(\nu-2)(\nu-1)$. This
is of course correct for small $qL$, i.e. near $\nu=2$, but we can find
a massive pole at large $qL$ where the above approximate
formula can not be applied any more. Then (\ref{braneGF}) and (\ref{prop}) lead
to the statement that we can see a massive four dimensional scalar and
the corrections from the KK modes to this massive scalar propagator. But it
should be noticed that this massive mode is not localized in the brane
since the wave function for this mode is not normalizable. This point
can be understood from the potential appearing in the bulk wave-equation
for the scalar in the fifth dimensional direction. The states
of $q^2>0$ would decay into large $z$ region outside the wall of 
the potential, so this state should
be regarded as a resonant state with a finite lifetime.
This is consistent with the results of \cite{DRT}, where lifetime
of the resonant state is explicitly given for small $M^2>0$.

%%%%%%%%%%%%  added A  %%%%%%%%%%%%%%
We can assure this point explicitly by rewriting $F(q,L,M)$ as
\beq
  F(q,L,M)={d-2\nu\over 2L}
     +q{J_{\nu}(qL)J_{\nu-1}(qL)-Y_{\nu}(qL)Y_{\nu-1}(qL)  
     +2i/(\pi qL)
        \over J_{\nu}(qL)^2+Y_{\nu}(qL)^2},
\eeq
and the resonant states are found as the zero points of the real
part of $F(q,L,M)$, where their imaginary parts are positive definite
by the formula 
$J_{\nu}(qL)Y_{\nu-1}(qL)-Y_{\nu}(qL)J_{\nu-1}(qL) = 2/(\pi qL)$.
This behaviour is seen at any value of $\nu$, and the positions of the
zero-points are slightly different near $q=0$. For typical values of
$\nu$, at $\nu=1.6, 2.0$ and $2.4$, $F(q,L,M)$ are shown in both regions
of $q>0$ and imaginary $q$ in Fig.~3 and Fig.~4, respectively.

%%%%%%%%%%%%%% Figs added %%%%%%%%%%%%%
\begin{figure}
\begin{center}
\includegraphics[width=7truecm]{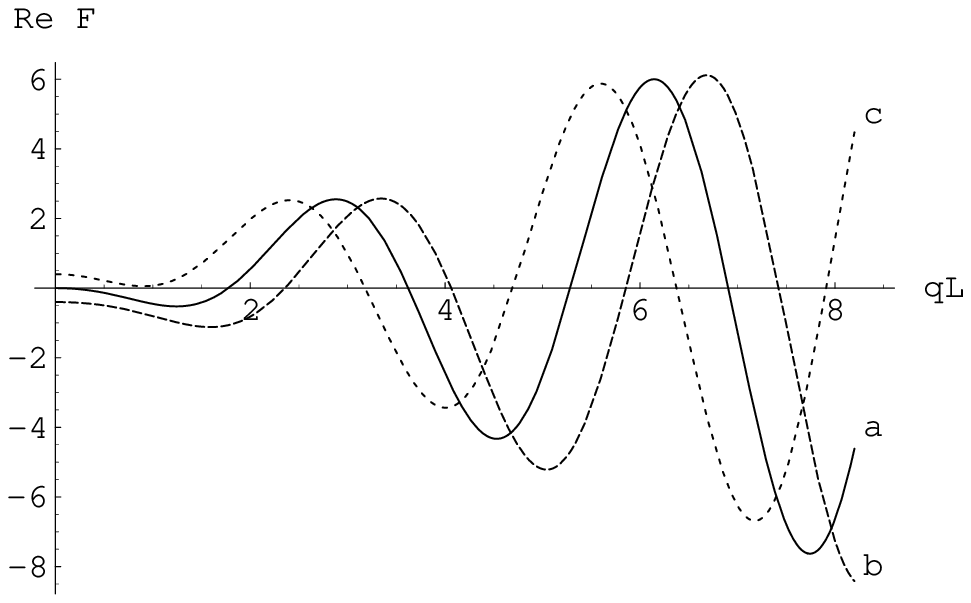}
\includegraphics[width=7truecm]{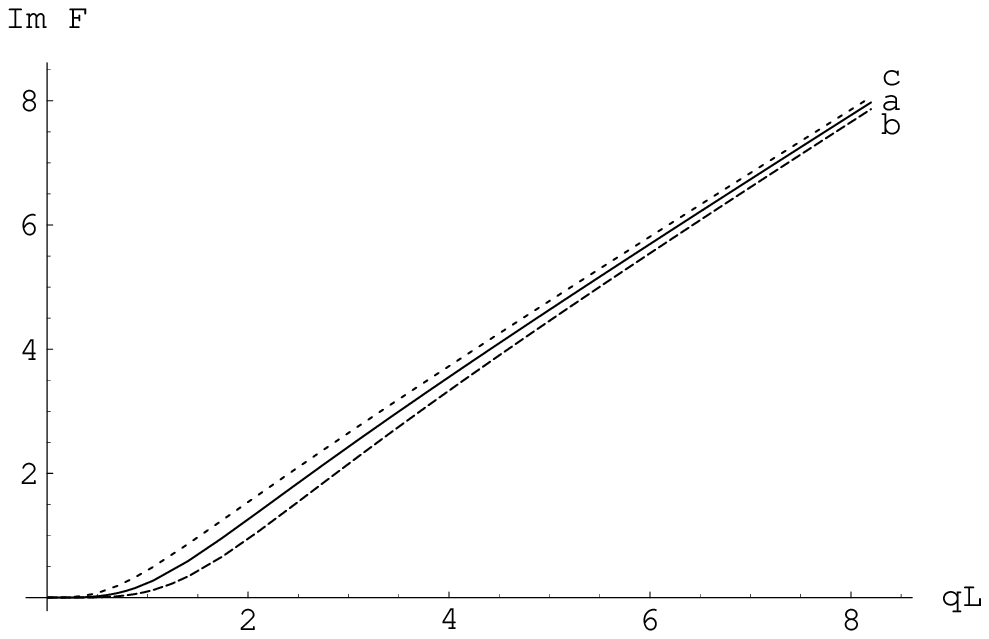}
\caption[]{$F(q,L,M)$ for real $q$. Curves $a, b$ and $c$
represent for $\nu=2.0, 1.6$ and 2.4 respectively.}
\end{center}
\end{figure}
%%%%%%%%%%%%%% Figs added %%%%%%%%%%%%%
%%%%%%%%%%%%%% Figs added %%%%%%%%%%%%%
\begin{figure}
\begin{center}
\includegraphics[width=12truecm]{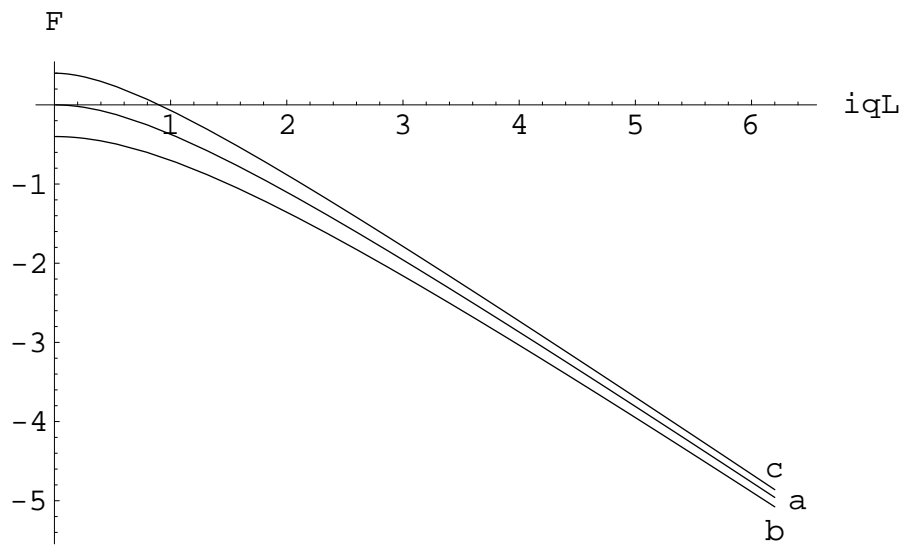}
\caption[]{$F(q,L,M)$ for imaginary $q$. Curves $a, b$ and $c$
represent for $\nu=2.0, 1.6$ and 2.4 respectively.}
\end{center}
\end{figure}
%%%%%%%%%%%%%% Figs added %%%%%%%%%%%%%

%%%%%%%%%%%%  added A %%%%%%%%%%%%%%

(iii) For $0<\nu<2$, we know that there is a localized state
with $q^2<0$ from the analysis of the wave-equation. In fact, we
can assure the same result also from the propagator given here.
The pole in the region $q^2<0$ is obtained as a zero point of $F(q,L,M)$
given in (\ref{prop}) by rewriting as $q\to iq$ in (\ref{prop}), i.e.
$F(iq,L,M)=0$, which is written as 
\beq
    (2+\nu)K_\nu(qL)=(qL)K_{\nu+1}(qL). 
\eeq
This is exactly the boundary condition (\ref{bound2}), 
required at the point of
brane in solving the wave equation. The pole point of $qL$ varies smoothly
from $0$ at $\nu=2$ to $1.5$ at $\nu=0$ as shown in the previous section. 
The approximate formula 
(\ref{appro}) is valid near $\nu=2$ also in this case. In any case,
we can see a tachyonic state and KK modes in the brane for $2>\nu\geq 0$. 

%%%%%%%%%%%%  added B  %%%%%%%%%%%%%%
The signal of the tachyon can be seen in the potential 
at the long distance as in the case of $\nu=2$,
where the massless pole provides the Coulomb potential $1/r$ at long
distance and the next leading $1/r^3$ from the KK-modes. In the case
of $\nu<2$, the leading potential comes from the tachyon pole, which
behaves as $\cos(|q|r)/r$, and the next leading coming from positive $p^2$
region is $1/r^3$. This is easily seen by expanding $F(q,L,M)$ at small
$q$ and performing the Fourier transformation of the leading terms.
So we can say that the dominant part of the potential at the 
long distance is given by the exchange of the tachyon.

%%%%%%%%%%%%  added B  %%%%%%%%%%%%%%

Then the brane would be unstable since the tachyon is not allowed
in the flat four dimensional brane.
%%%%%%%%%%%%%%%%%%%%%%5

We can summarize the above results from the viewpoint of the
exchange of $\phi$ fields between two sources.
For $\nu=2$, we can see that at large
distances on the brane, $|x-x'|\gg L$, the zero mode piece dominates and we
reproduce the standard effective action for $d$-dimensional scalar
exchange, plus sub-leading corrections from the Kaluza-Klein part.
This picture is not true for $\nu>2$, and we can not see any Coulomb like
behaviour. For $\nu<2$, the tachyon which is allowed in the bulk $AdS_5$
penetrates into the brane as a tachyon in the four dimensional brane,
%%%%%%%%%%%%  added C  %%%%%%%%%%%%%%
and it provides the dominant part of the potential between two charges
at long distance.
%%%%%%%%%%%%  added C  %%%%%%%%%%%%%%
This
leads to the instability of the RS brane with flat four-dimensional metric.
A simple way to remove this instability is to add the following quadratic
term of the scalar $\Phi$ to the brane action $S_{\rm bl}$,
\beq
   \int d^4x \sqrt{-g}{1\over 2}{d-2\nu \over L}\Phi^2 . \label{new1}
\eeq
Since this term could cancel out the first term of $F(q,L,M)$
given in (\ref{prop}), which represents the effective quadratic
term of the scalar $\Phi$ in the brane, then we obtain the same
propagator with the one of $M=0$. Then we can see the localized zero
mode for wide range of $\nu$, i.e. for $\nu>1$. 

If we adopt the term (\ref{new1}), the boundary condition of the Green function
is changed. By taking into account this term with a coefficient
independent on the scalar mass $M$,
the problem of the localization of a scalar field was also discussed in 
\cite{MPi} from
a different viewpoint and in a different region of its mass with our analysis.
At present, we
can't say anything definitely about the meaning of this new term.

%%%%%%%%%%%%  added D  %%%%%%%%%%%%%%
More probable resolution of this instability will be to solve the equations
of motion for the system of gravity and a scalar with a localized brane-action.
It would be a challenging subject to find a solution of Randall-Sundrum
type brane from a supersymmetric model induced from the super-string theory.
And the action would contain a non-trivial potential and the scalar
field would be solved as a non-trivial configuration forming the background
\cite{DeW}.
The metric might be solved in the form
\beq
 ds^2= e^{A(y)}\eta_{\mu\nu}dx^{\mu}dx^{\nu}
           +dy^2  \, \label{metrica2}
\eeq
and the cosmological constant on the brane can be taken as zero.

In this case, the instability pointed out here would be evaded by
the background configuration which control the equations of the
fluctuation of the scalar and gravitation.

%%%%%%%%%%%%  added D  %%%%%%%%%%%%%%

In the next section, we discuss how this instability would be resolved 
from the viewpoint of AdS/CFT correspondence. It would give a clue to
consider the meaning of the above newly added term
%%%%%%%%%%%%  added E  %%%%%%%%%%%%%%
and the string induced bulk-action with a scalar.
%%%%%%%%%%%%  added E  %%%%%%%%%%%%%%

%%%%%%%%%%%%%%%%  Duality below %%%%%%%%%%%%%%%%%%%%%%%%%%%%%%%%%%

\section{ AdS/CFT correspondence }

The gravitational theory in the background of $AdS_5$ is dual to the 4d
conformal field theory with a cutoff at the position of the RS brane, so
we can replace the five dimensional bulk part by a CFT defined as
\beq
  S_{\rm CFT}=\int d^4x( L_{\rm CFT}+\Sigma_i\lambda_iO^i), \label{cft}
\eeq
where $O^i$ are the composite operators of the fields contained 
in $L_{\rm CFT}$
and $\lambda_i$ are their corresponding
sources which are given as the boundary values
of the bulk fields at the brane position.
Here we should consider both the gravitation and the scalar
on the brane.

Then the effective brane action could
be obtained by adding the regularized part of 
intrinsic curvature \cite{HeSk,BaKr,11} 
and the localized scalar part, $L_{\phi}$, as follows
\beq
    S_{\rm b2} = \int dx^4\sqrt{-g}\Bigg\{L_{\rm brane}+L_{\phi}
              -({\tau\over 2}+b_0)
     -b_2R-b_4R_2\Bigg\}, \label{baction2}
\eeq
where $b_0=-(6/L)/(2\kappa^2)$, $b_2=-(L/2)/(2\kappa^2)$, 
$b_4=2L^3/(2\kappa^2)$ and
\beq
  R_2 = -{1\over 8} R_{\mu\nu} R^{\mu\nu} + {1\over24} R^2\ .
\eeq
Since $\tau=(12/L)/(2\kappa^2)$, the cosmological constant on the brane, 
$\tau/2+b_0$, is zero. This is consistent with the Poincare invariant solution.
The four dimensional curvature term reflects the localized zero mode of gravity
and the curvature squared terms are the correction to the gravity part. 
%%%%%%%%%

%%%%%%%
The corrections to $S_{\rm b2}$ would be
obtained by the CFT part. It is seen by integrating the fields contained 
in $L_{\rm CFT}$. In the case of the graviton, this correction has been 
explicitly shown within one-loop approximation \cite{DL}, 
and the correspondence of this
correction and the sum up of the KK modes has been assured.

As for the scalar field, it is difficult to give the explicit form
of the composite operator and to separate
the propagator obtained in the previous section into two parts,
the bounded part and KK modes for general $\nu$. 
Then we don't try to see the correspondence for the quadratic term of $\phi$
here as shown for the graviton. Instead, we use the AdS/CFT correspondence
to see the effective action of the scalar $\phi$ for a calculable case.
As in \cite{DL}, we consider ${\cal{N}}=4$ supersymmetric Yang-Mills theory 
as the CFT which lives on the boundary of the $AdS_5$.
The conformal dimension of the operator is given by $2+\nu$, so the operator
is relevant for $\nu<2$. Here we consider the case of $\nu=0$.

For $\nu=0$,
we can consider the mass term of the adjoint scalar fields,
which are denoted by $(B_i)^2$ ($i=1\sim 6N^2$),
of $SU(N)$ Yang-Mills theory as the corresponding composite operator
since it is gauge invariant
and has the conformal dimension two. The mass term of the vector would break
gauge symmetry, so we do not consider it.

By assuming the availability of the AdS/CFT correspondence,
we consider the combined action, $S_{\rm CFT}+S_{\rm b2}$,
to obtain the effective potential of $\phi$ with its higher powers.
It is easy to obtain the one-loop correction by integrating the fields
in CFT. In the case of $\nu=0$, we obtain
\beq
 V_0(\phi)=-m_0^2{\phi}^2+{6N^2\over 64\pi^2}\phi^2\ln{\phi\over \mu^2},
          \label{effpo}
\eeq
where $m_0$ denotes the tachyon mass obtained above for $\nu=0$ and
$\mu$ is some mass scale. The first term is included in $L_{\phi}$
in (\ref{baction2}), and $6N^2$ represents the degrees of freedom
of the adjoint scalar in CFT. Since $L_{\phi}$ should be obtained from
(\ref{norm1}) and (\ref{norm}), then the notation $\tilde{\phi}$ 
should be used instead $\phi$. And
$\phi$ of the second term should be written by
multiplying $\hat{\psi}(0)$ as $\hat{\psi}(0)\phi$, which represents $\Phi$ on
the brane. These points are important when we consider 
the relative ratio of the two terms and the strength of
the coupling of the boundary field and composite operator of CFT.
But they are not essential hereafter, and we abbreviate its notation
as $\phi$.

Further, we should notice that the second term is
proportional to $\phi^2$ not to $\phi^4$ since $\phi$ couples to
the mass term of the CFT scalar (denoted by $O_0$) as $\phi O_0$ not
as $\phi^2O_0$ due to its conformal dimension two.
Then the above potential
can be rewritten as
\beq
 V_0(\phi)={6N^2\over 64\pi^2}\phi^2\ln{\phi\over \hat{\mu}^2},
          \label{effpo2}
\eeq
where new scale factor $\hat{\mu}$ is defined as
${6N^2\over 64\pi^2}\ln{\mu^2\over \hat{\mu}^2}=-m_0^2$. 
Here, we notice that this potential has the same form, except for
its sign, with the
one obtained from open string field theory \cite{GS,KMM,MZ} for the
tachyon field. This seems curious since there is no definite reason to 
identify RS brane as D-brane.

%%%%%%%%%%%%%%%% Modify 1 %%%%%%%%%%%%%%%%%%%
If this approximation is available, the above result (\ref{effpo2})
implies that the scalar field should be shifted to see its stable
spectrum from the tachyonic point $\phi=0$ to the minimum point
of $V_0(\phi)$. 
%%%%%%%%%%%%%%%% Modify 1 %%%%%%%%%%%%%%%%%%%
However, the value of $V_0(\phi)$ at its minimum is negative, so the
flat metric in the brane is still unstable if there is no extra term to
cancel this new cosmological constant. 
This implies
that we must return to the point where we solve the bulk equations
to obtain the background metric, $AdS_5$, by taking into account of
the non-trivial configuration of the scalar fields with negative
mass squared. It would be expected in this case that the stable
RS brane in the $AdS_5$ background would be obtained by a shift
of the scalar. Its fluctuation could be observed 
as a massive particle around a bottom of its 
effective potential not a tachyon around the top. A simple example would be
shown in a separate paper. 

%%%%%%%%% added %%%%%

%%%%%%%%%%%%%%%%%%%%%

It is an open problem to extend
this story to the other value of $\nu>0$ or to the other dimensional $AdS$.\footnote{The idea of $AdS_d$/CFT correspondence for RS brane has been
discussed in \cite{ANO} for the case of $d=3\sim 5$.
Also in \cite{NOZ}, we can see a related discussion from a 
different formalism.} 
Here we obtain a resultant potential term, $\phi^2\ln{\phi\over \mu^2}$,
for the tachyonic scalar $\phi$. This form is slightly different from the one
considered in the previous section, (\ref{new1}). The problem related to
this point would be discussed in the preparing paper.

%%%%%%%%%%%%%%%%%  two branes %%%%%%%%%%%%%%%%%%%%%
\section{Two branes}

Here we consider the model of two branes proposed by Randall and
Sundrum \cite{RS1}.
In this case, the distance between two branes in the direction of the fifth
coordinate is fixed as finite. So the concept of the localization can't be 
considered in a strict sense. All modes are confined in the region between the 
two branes, and $m$ would take discrete values instead.
However, the localization of the zero gravitational mode would
be important in a sense of the wave-function localized near one brane if
our world is not five dimensional one.

A problem of this model is to determine the distance at some definite
value as a stable point of the brane system. Several people proposed
ideas of the resolutions
for this issue by considering the bulk fields \cite{GW,LS}. In 
\cite{GW}, a scalar field has 
been introduced with special potentials on the branes in order to
fix its boundary values. It can be seen from the effective potential 
of this scalar zero-mode\footnote{Here we use ``zero-mode'' in the 
sense that the eigenvalue of the four-dimensional Laplacian is 
zero in the equation.} that there is a stable point at a finite 
distance between two branes.  
While the scalar-potential on the branes
plays an important role to solve the problem, its origin is obscure.

Here we do not give the answers to these problems. We comment on this
problem by considering the scalar without any potential in the branes.
The scalar considered in the previous sections is this type of scalar.
The strategy is to see the effective potential by integrating the action
of the scalar part by substituting the solution of the wave equation
in the AdS background with two branes on the boundaries. The equation to
be solved is given by (\ref{scalar}) $\sim$ (\ref{poten}) and the boundary
conditions, (\ref{bound}) at $z=0$ and 
\beq
 \partial_z\hat{\psi}|_{z_c}=-{3\over 2L(z_c/L+1)}\hat{\psi}|_{z_c},
        \label{boundzc}
\eeq
at $z=z_c$ where the second brane exists.

We give some comments on the solution in this case. The general solution
of $\hat{\psi}$ can be written by the linear combination of two independent
Bessel functions. The two boundary conditions (\ref{bound}) and (\ref{boundzc})
determine a relation of two coefficients of the Bessel functions and
the four dimensional mass-eigenvalue of the allowable state. 
The eigenvalues are discrete
in this case. So it would be difficult to sum up these discrete states to see
the effective potential totally
as a function of the distance between two branes.
However we can see it easily in our case.

The equations to be solved are rewritten as the equations of $\chi(y)$,
where it is defined as $\Phi=\phi(x)\chi(y)$. Then
the above boundary conditions are written as,
\beq
 \partial_y\chi|_{y_1}=\partial_y\chi|_{y_2}=0,   \label{boundy}
\eeq
where $y_1(=0)$ and $y_2$ are the positions of the two branes. On the other
hand, the scalar part of the action can be written as follows by using its 
equation of motion in the AdS background;
\bea
  S_{\Phi}^{\rm classical} &=& -{1\over 2}\int_{y_1}^{y_2} dy\int d^4x\, 
    \sqrt{-G}(G^{AB}\partial_A\Phi\partial_A\Phi + M^2\Phi^2) \cr
%  &=& -{1\over 2}\int_{y_1}^{y_2} dy\int d^4x\,
%      \partial_A\left(\sqrt{-G}G^{AB}\Phi\partial_B\Phi\right) \cr
  &=& -{1\over 2}\int_{y_1}^{y_2} dy\int d^4x\,
       \partial_\mu\left(\sqrt{-G}G^{\mu\nu}\Phi\partial_\nu\Phi\right) \cr
  &&  -{1\over 2}\int d^4x\,\left(\sqrt{-G}G^{yy}\Phi\partial_y\Phi\right) 
                           \Bigr|_{y_1}^{y_2}. 
\eea
The first term vanishes since it is total derivative and the second term 
vanishes due to the boundary conditions (\ref{boundy}).  
Then we can see $S_{\Phi}^{\rm classical}=0$ so that the scalar 
considered here has nothing to do with the stabilization of two 
branes at some finite distance. 

As in the case of \cite{GW}, if some
boundary terms of the scalar are added to the brane actions, then
the boundary conditions are changed. In this case, 
$S_{\Phi}^{\rm classical}\neq 0$. So
we must solve the mass spectrum allowed between two branes and they are summed
up to obtain the effective potential as a function of the distance between branes. We are now
preparing the paper of this issue.

%%%%%%%%%%%%%%%%  Conclusion %%%%%%%%%%%%%%%%%%
\section{Concluding remarks}

When we consider some supersymmetric five dimensional theory to construct the 
brane as a soliton or a domain wall, we must aware about the contents of the
theory even if it was not used to solve the equations responsible for the 
background. It would be important to study the behaviours of various fields
in the background obtained as a classical solution.
Here we have examined the scalar field in 
the AdS background, in which the Randall-Sundrum three-brane(s) is
(are) embedded. The mass-squared of the scalar is extended
to the allowable negative value in the $AdS_5$ background.
We found a localization-mode of the scalar, which has negative mass-squared
in the bulk, on the brane, and this localized mode is also tachyonic
in the four dimensional brane-world.
%%%%%%%%%%%%  Modified below  %%%%%%%%%%%%%%%%%%%%%%%%
This fact implies that the three-brane 
with flat space would be unstable since a tachyonic scalar is living
in this brane world. Other continuous tachyonic-modes would not cause
any instability of the three brane since they should be considered as 
the bulk tachyonic scalar being allowed in $AdS_5$.

It would be possible to consider several ways to evade this problem 
of the instability of the brane. (i) One way is to improve the brane-action
by adding a boundary potential of the scalar such that it cancels the
tachyonic mass of the localized state. As a result, we could find a
massless localized scalar on the brane. But this procedure seems to be
artificial since the boundary potential is added by hand.
(ii) Second way is to solve the bulk equations by assuming a non-trivial
configuration for the tachyonic scalar, and we try to find a solution
which forbids the tachyonic localized state. Or we should obtain a brane
solution which allows tachyonic localized-scalar. In this approach, it
would be necessary to know the fully corrected,
complicated potential for the scalar.
(iii) As a third possibility to resolve this problem, we have examined
the effects of the bulk on the brane from the viewpoint of AdS/CFT
correspondence. For $\nu=0$, we find an interesting scalar potential
which can be related to this problem, but we could not find a general
clue to resolving the problem in this direction. To get some definite
conclusion in this approach, we should examine the properties of many
more bulk-operators. 

We will discuss on these points in the future paper.
We have also examined the role of the tachyonic scalar
in the case of two branes, but we find that
this scalar gives no effect to the problem of
determining the stable point between two branes.

\vspace{.5cm}

\noindent {\bf Acknowledgement:} We wish to thank Drs. S. Dubovsky and
S. Shatashvili for giving us useful informations and references.


\vspace{.3cm}


\begin{thebibliography}{99}

\bibitem{M1} J.~Maldacena, Adv. Theor. Math. Phys. {\bf 2} (1998) 231, 
        ({\tt hep-th/9711200}).
\bibitem{GKP1} S.S. Gubser, I.R. Klebanov, and A.M. Polyakov Phys. Lett.
        {\bf B 428} (1998) 105, ({\tt hep-th/9802109}).
\bibitem{W1} E.~Witten, Adv. Theor. Math. Phys. {\bf 2} (1998) 253, 
        ({\tt hep-th/9802150}).
\bibitem{Poly1} A.M. Polyakov, Int. J. Mod. Phys. {\bf A14} (1999) 645,
        ({\tt hep-th/9809057}).
\bibitem{RS1} L. Randall and R. Sundrum, Phys. Rev. Lett. {\bf 83} (1999) 3370,
        ({\tt hep-ph/9905221}).
\bibitem{RS2} L. Randall and R. Sundrum, Phys. Rev. Lett. {\bf 83} (1999) 4690,
        ({\tt hep-th/9906064}).
\bibitem{ADD} N. Arkani-Hamed, S. Dimopoulos, and G. Dvali, 
        Phys. Lett. {\bf B429} (1998) 263, ({\tt hep-ph/9803315}).
\bibitem{AADD} L. Antoniadis, N. Arkani-Hamed, S. Dimopoulos and G. Dvali, 
        Phys. Lett. {\bf B436} (1998) 257, ({\tt hep-ph/9804398}).
\bibitem{Ant} L. Antoniadis, Phys. Lett. {\bf B246} (1990) 377.
\bibitem{ArHa} N. Arkani-Hamed, S. Dimopoulos, N. Kaloper and R. Sundrum, 
        Phys. Lett. {\bf B480} (2000) 193, ({\tt hep-th/0001197}).
\bibitem{KaSc} S. Kachru, M. Schulz and E. Silverstein, Phys. Rev. {\bf D62} 
        (2000) 045021, ({\tt hep-th/0001206}).
\bibitem{11} S.B. Giddings, E. Katz and L. Randall, JHEP {\bf 03} (2000) 023, 
        ({\tt hep-th/0002091}).
\bibitem{DRT} S.L. Dubovski, V.A. Rubakov and P.G. Tinyakov, Phys. Rev. 
         {\bf D62} (2000) 105011, ({\tt hep-th/0006046}).
\bibitem{DL} M.J. Duff and J. T. Liu, Phys. Rev. Lett. {\bf 85} (2000) 2052,
        ({\tt hep-th/0003237}).
\bibitem{GS} A.A. Gerasimov and S.L. Shatashvili,
         JHEP {\bf 034} (2000) 0010, ({\tt hep-th/009103})
\bibitem{KMM} D. Kutasov, M. Marino and G. Moore,
         JHEP {\bf 045} (2000) 0010, ({\tt hep-th/009148})
\bibitem{MZ} J.A. Minahan and B. Zwiebach,
         JHEP {\bf 029} (2000) 09, ({\tt hep-th/000823})
\bibitem{PPN} M. Pernici, K. Pilch and P. van Nuiuwenhuizen, 
         Nucl. Phys. {\bf B259} (1985) 460.
\bibitem{BC} K. Behrndt and M. Cvetic, Phys. Rev. {\bf D61} (2000) 101901,
        ({\tt hep-th/0001159}).
\bibitem{Be} K. Behrndt and M. Cvetic, Phys. Lett. {\bf B475} (2000) 253,
        ({\tt hep-th/9909058}).
\bibitem{DuLi} M.J. Duff, J.T. Liu and K.S. Stelle, ``A supersymmetric Type IIB 
        Randall-Sundrum realization'', {\tt hep-th/0007120}.
\bibitem{Bre} P. Breitenlohner and D.Z. Freedman, Ann. Phys. {\bf 144}
 (1982) 249; Phys. Lett. {\bf B115} (1982) 197.
\bibitem{MPi} M. Mintchev and L. Pilo, ``Localization of Quantum Fields on 
        Branes'', {\tt hep-th} {\tt /0007002}.
\bibitem{DeW} O. DeWolfe, D.Z. Freedman, S.S. Gubser and A. Karch, 
        Phys. Rev. {\bf D62} (2000) 046008, ({\tt hep-th/9909134}).
\bibitem{HeSk} M. Henningson and K. Skenderis, 
        JHEP {\bf 07} (1998) 023, ({\tt hep-th/9806087}).
\bibitem{BaKr} V. Balasubramanian and P. Kraus, Commun. Math. Phys. 
        {\bf 208} (1999) 413, ({\tt hep-th/9902121}).
\bibitem{ANO} L. Anchordoqui, C. Nunez and K. Olsen, 
        JHEP {\bf 10} (2000) 050, ({\tt hep-th/0007064}).
\bibitem{NOZ} S.Nojiri, S. Odintsov and S. Zerbini, Phys. Rev. {\bf D62},
          (2000) 064006, ({\tt hep-th/0001192}).
\bibitem{GW} W.D.~Goldberger and M.B.~Wise, Phys. Rev. Lett. {\bf 83} (1999) 
     4922, ({\tt hep-ph} {\tt /9907447}).
\bibitem{LS} M. A. Luty and R. Sundrum, Phys. Rev. {\bf D62} (2000) 
        035008, ({\tt hep-th/9910202}).

%%%%%%%%%%%%%%%%%%%%%%%%%%%%%%%%%%%%%%%%%%%%%%%%%%%%%%%%%%%%%%%%%%%%%%%%%%
\end{thebibliography}
\end{document}